\documentclass[aps,prl,twocolumn,reprint,superscriptaddress,nofootinbib,
tightenlines,preprintnumbers,showkeys,floatfix]{revtex4-2}

\usepackage[utf8]{inputenc}
\usepackage[english]{babel}
\usepackage{amssymb,amsthm,amsmath,amstext,amsbsy,amsopn,mathrsfs}
\usepackage{bbm}
\usepackage{nicefrac}
\usepackage{slashed}
\usepackage{graphicx}
\usepackage{hyperref}
\usepackage{leftidx}
\usepackage{environ}
\usepackage{mathtools}
\usepackage{xspace}
\usepackage{array}
\usepackage{xcolor}
\usepackage{isotope}
\usepackage{suffix}
\usepackage{orcidlink}
\usepackage{booktabs,colortbl}
\usepackage[caption=false]{subfig}
\usepackage{braket}

\newcommand{\T}{\mathbf T}

\begin{document}

\title{Constructing Effective Interactions via Projection-Based Inversion}

\author{Hang Yu\,\orcidlink{0000-0001-6860-5960}}
\email{yhang@nucl.ph.tsukuba.ac.jp}
\affiliation{Center for Computational Sciences, University of Tsukuba,
Tsukuba, Ibaraki 305-8577, Japan}

\author{Serdar Elhatisari}
\email{selhatisari@gmail.com}
\affiliation{Faculty of Natural Sciences and Engineering,
Gaziantep Islam Science and Technology University, Gaziantep 27010, Turkey.}

\author{Sebastian König}
\email{skoenig@ncsu.edu}
\affiliation{Department of Physics and Astronomy,
North Carolina State University,
Raleigh, NC 27695, USA}

\author{Dean Lee}
\email{leed@frib.msu.edu}
\affiliation{Facility for Rare Isotope Beams \& Department of Physics
and Astronomy, Michigan State University, MI 48824, USA}

\author{Yuan-Zhuo Ma}
\email{mayu@frib.msu.edu}
\affiliation{Facility for Rare Isotope Beams \& Department of Physics
and Astronomy, Michigan State University, MI 48824, USA}

\author{Takayuki Miyagi}
\email{miyagi@nucl.ph.tsukuba.ac.jp}
\affiliation{Center for Computational Sciences, University of Tsukuba,
Tsukuba, Ibaraki 305-8577, Japan}

\begin{abstract}

We present a numerical prescription for extracting continuum scattering
information from discrete spectra by constraining effective interactions
inspired by effective field theory (EFT).
Using a Multiparameter Eigenvalue Problem (MEP) emulator, we map energies to a
sum of contact potentials by recasting the inverse problem as a linear
eigenvalue equation.
Because our method determines the effective interaction rather than the
scattering amplitude, it can handle non-perturbative Coulomb interactions and
different types of truncated Hilbert spaces without analytic quantization
conditions. It therefore allows standard bound-state codes to be
used for scattering calculations without modification.
We validate this prescription
across multiple \textit{ab initio} frameworks using neutron-alpha scattering,
alpha-alpha scattering with full Coulomb, and a prediction of
proton-\isotope[14]{O} resonances.
\end{abstract}

\maketitle


\paragraph{Introduction.}
\textit{Ab initio} nuclear physics has made tremendous progress over the past
two decades, fueled by the development of interactions based on or inspired by
effective field theories~\cite{Hammer:2019poc,LENPIC:2022cyu,Machleidt:2024bwl} 
and by advances in computational techniques for solving the nuclear many-body
problem (see, e.g., Refs.~\cite{Hergert2020,Lee2025,Launey:2025qdd}).
While this progress also includes calculations of nuclear reactions from first
principles, methods for solving bound-state-like problems are
generally still far ahead of \textit{ab initio} techniques for calculating nuclear
reactions~\cite{Johnson2020}, and calculations become particularly challenging near
the edges of nuclear stability~\cite{Bazin:2022gfo,Crawford:2023txq}.
One reason for this is that the complexity of continuum calculations rises much
more steeply with the number of nucleons than it does for bound-state problems.
Significant effort has to be spent on merely constructing the asymptotic states
out of nucleons, and as the energy increases, a full \textit{ab initio}
treatment requires the inclusion of more and more channels, many of which are
coupled by the interaction~\cite{Navratil:2016ycn,Hergert2020}.

While to some extent the problem can be tackled with a combination of physical
insight and brute computational force (see, for example,
Refs.~\cite{Deltuva2008,Leidemann2013,Navratil:2016ycn,Navratil:2022lvq,%
Elhatisari:2015iga,Navratil:2016ycn,Navratil:2026hbo,Sarkar:2026wev} and further
references therein), a more economical approach is to map an initial fully
interacting many-body system onto a system of clusters described by effective
interactions.
If this mapping is performed in a rigorous and systematic manner, it becomes
possible to perform calculations at greatly reduced cost while maintaining the
\textit{ab initio} nature of the overall approach.

Effective field theories (EFTs)~\cite{Hammer:2019poc} provide the general
framework for such a development, and in particular Halo/Cluster EFT has been
conceived~\cite{Bertulani:2002sz} as an approach that can explicitly utilize
scale separations in certain nuclei to construct a systematically improvable
description in terms of a series of contact interactions between a ``core''
and one or more ``valence nucleons'' (Halo EFT), between multiple clusters
(Cluster EFT), or, in principle, a combination of both.
The coupling strengths of these interactions, generally referred to as
``low-energy constants (LECs),'' need to be determined from experimental data,
or from some more microscopic calculation.

In this work, we pursue a simplified formulation of this latter approach that
sets aside certain technical aspects of the EFT, particularly questions of
rigorous renormalization (see Ref.~\cite{Hammer:2019poc} for a detailed
discussion), and instead focuses on a novel numerical technique for performing
the theory mapping process.
More precisely, we map the microscopic theory onto an EFT-inspired model
(describe in more detail below), and we present a highly efficient method
for constraining such effective interactions through a matching of discrete
energy levels between microscopic (``high-resolution'') and macroscopic
(``low-resolution'') representations of a given system.

The system's discrete energy spectrum may arise either explicitly from a
formulation within a finite volume (such as a cubic box or a harmonic trap), or
implicitly from working in a truncated harmonic oscillator (HO) basis (without
explicit trapping potential).
Either way, the spatial confinement discretizes the continuum spectrum and
encodes scattering information in how the energy levels shift when the size of
the volume (or, more generally, the basis truncation) is varied.
Within the context of Lattice Quantum Chromodynamics (Lattice QCD or LQCD),
the formalism for decoding these shifts has been pioneered by
\textcite{Luscher:1986pf,Luscher:1990ux} for periodic finite boxes.
Recent work in this context has increasingly focused on connecting
the finite-volume spectrum directly to the LECs of an EFT (Pionless EFT, in
particular, see the review by \textcite{Hammer:2019poc})~\cite{Briceno:2013lba,
Barnea:2013uqa, Hall:2013qba, Hall:2014uca, Wu:2014vma, Briceno:2017max, 
Detmold:2021oro, Meng:2021uhz, Kirscher:2015yda, Yaron:2022rmb}.
\textcite{Busch:1998cey} formulated a quantization condition for harmonic traps,
while Ref.~\cite{Shirokov:2003kk} developed the formalism for truncated HO
spaces.
The latter is particularly relevant in the nuclear physics context because a
variety of \textit{ab initio} methods work with truncated HO spaces as
the underlying basis to formulate the many-body problem in.

In this work, we develop numerical machinery that encompasses all of these
approaches, constructing a framework that encodes the relevant
physics properties in the coupling constants of a generic effective interaction.
Specifically, we use the Multiparameter Eigenvalue Problem (MEP) emulator
method~\cite{Yu:2025mep} to ``reshuffle'' the Schrödinger equation: instead of
solving for energies given fixed interaction parameters (the forward problem), our
method treats the \emph{coupling constants} as solutions of a generalized 
eigenvalue problem, with the microscopic energies as input constraints.
This procedure is distinctly different from a standard optimization (fitting)
problem.
Through subspace projection, we achieve an algebraic inversion of the
forward problem that maps a \emph{domain} of spectral information to the 
effective interaction.

Because our MEP method works at the level of the effective interactions
rather than the scattering amplitude, it does not rely on the analytic
properties of the amplitude that constrain L\"{u}scher's original formalism.
For instance, the Coulomb interaction, the infinite range of which often
requires perturbative and/or numerically complicated modifications to
analytical quantization conditions~\cite{Davoudi:2014qua,Beane:2014qha,%
Borsanyi:2014jba,Stellin:2020gst,Guo:2021qfu,Guo:2021lhz,Yu:2022nzm}, is
handled non-perturbatively by simply including it in the Hamiltonian.
Therefore, no analytic modification is needed.
Other amplitude-level difficulties, including coupled channels~\cite{He2005,%
Guo2013,Wu:2014vma,Briceno:2017max,HadronSpectrum2020} and three-body
quantization conditions~\cite{Polejaeva:2012ut, Hansen:2015zga,%
Hammer:2017uqm,Hammer:2017kms,Muller:2021uur}, are similarly avoided since
the method never constructs the scattering amplitude as an intermediate
object.
Moreover, because the method extracts interactions directly from
truncation-dependent spectra, bound-state frameworks such as the No-Core
Shell Model (NCSM)~\cite{Barrett2013}, Nuclear Lattice EFT
(NLEFT)~\cite{Lee2025}, and the In-Medium Similarity Renormalization
Group (IMSRG)~\cite{Hergert2017,Stroberg:2019mxo} can be used for
scattering caclculations without modification.
The only inputs required are energies at different truncations -- a task
that is already standard in convergence tests of many-body methods.


\paragraph{Methods.}
We consider a clustered system described by an effective few-body Hamiltonian
in a truncated space:
\begin{equation}
    H_{\rm trunc}(\T)
    = \underbrace{K(\T) + V_C(\T)}_{H_0(\T)} + V_{\rm eff}(\T)\,,
\end{equation}
where $K$ is the kinetic term, $V_C$ is the Coulomb interaction, and
$V_{\text{eff}}$ is the short-range effective interaction between clusters,
to be determined by our MEP formalism from \textit{ab initio} input.
In this letter, we limit our discussion to the single-channel two-body, such
that $ V_{\text{eff}}$ involves two-body interactions only.
We use $(\T)$ to indicate the truncation of this Hamiltonian, which is chosen
to match the truncation of the many-body space on the microscopic
(\textit{ab initio}) end.
In the most general sense, we note that this truncation $\T= (T_l, T_h)$
includes both an infrared (IR) scale $T_l$ and an ultraviolet (UV) scale $T_h$.
For example, on a periodic lattice, the IR truncation is $T_l = L$, the total
size of the box, and the UV truncation is $T_h = a_{\text{latt}}$, the lattice
spacing.
For an HO basis, the IR truncation is $T_l = L_{\text{eff}}
= \sqrt{(2e_{\text{max}} + 7)} b$~\cite{Furnstahl:2013vda} and the UV
truncation is $T_h =b/ \sqrt{(2e_{\text{max}} + 7)} $~\cite{Konig:2014hma} 
(with $b^2 = \frac{\hbar }{\mu\Omega}$).
Both are defined uniquely from the HO frequency $\hbar\Omega$, the basis size
specified by either $e_{\text{max}}$ or $N_{\text{max}}$, and the reduced mass
$\mu$ of the clusters.

Inspired by Halo and Cluster EFT~\cite{Capel:2018kss,Capel:2021ejr,%
Hammer:2022lhx}, we expand $V_{\text{eff}}$ in a basis of operators.
This allows for numerically implementing power counting schemes.
We use coordinate-space bases and numerical regulators to match the
truncated bases in \textit{ab initio} methods.
We employ two formulations.
The first is a series of local Gaussians as a function of the relative
distance $r$:
\begin{equation}
    V_{\text{eff,local}} =  C_0 e^{-\Lambda^2 r^2} 
    + C_2 \left(\Lambda^2 r^2 e^{-\Lambda^2 r^2}\right) + \cdots \,,
    \label{eq:expansion-local}
\end{equation}
where $C_{2n}$ are the couplings to be determined and $\Lambda$ is the
regularization momentum scale.
To test different schemes, we also use a non-local separable potential, 
\begin{multline}
    V_{\text{eff,sep} }=  C_0 \ket{d} \bra{d}
    + \\
    C_2 \int\mathrm{d} \mathbf{r} \, {e}^{{-}\Lambda^2 r^2} \Big(
     \ket{d} \bra{ \mathbf{r}} + \ket{\mathbf{r}} \bra{d}
    \Big)+ \cdots \,,
\label{eq:expansion-sep}
\end{multline}
with an additional vector $\ket{d}$ inspired by the
finite-volume Hamiltonian method~\cite{Hall:2013qba,%
Hall:2014uca,Wu:2014vma,Yk:2025gzg}; see also Ref.~\cite{Briceno:2013lba}
for an alternative formulation with a dimer field. 
We find that both formulations yield similar results within truncation errors.

The core of our method is to determine the subspace-projected inverse mapping
from energy levels $\{E(\T_j)\}$ obtained at various truncations $\T_j$ to
the couplings $\{C_{2n}\}$.
This leads to the coupled equations:
\begin{multline}
  \left( H_{0}(\T_j) - E(\T_j) + \sum_{i=1}^{n_{\text{par}}} 
  C_{2(i-1)} V_{i}(\T_j) \right) \ket{\psi(\T_j)} 
  = 0\,,\\~~~ 1\leq j\leq m \,.
\label{eq:Main}
\end{multline}
This is an MEP where the coupling constants $C_{2n}$ are the eigenvalues,
the energies $E(\T_j)$ are input parameters, and the $V_i$ are
the $n_{\text{par}}$ operators in the truncated effective interaction.
Exact determinacy requires $m = n_{\text{par}}$, and we vary the selection
of $\T_j$ to validate consistency.
The system is rewritten as a generalized eigenvalue problem for each $i$:
\begin{equation}
    \left(\mathbf K_i\left(E(\T_1), E(\T_2),\ldots \right) 
    +  C_{2(i-1)} \mathbf K_0 \right) \ket{\mathbf{\Psi}_{\otimes}} = 0\,,
    \label{eq:GEM}
\end{equation}
where $\ket{\mathbf{\Psi}_{\otimes}} = \bigotimes_j \ket{\psi(\T_j)}$ is
the Kronecker product of wave functions across $\T_j$, and $\mathbf{K}_i$
are the MEP operator determinants~\cite{Yu:2025mep}.
We solve this high-dimensional problem using a projection-based
reduced-order model built on eigenvector continuation~\cite{Frame:2017fah, Duguet:2023wuh}.
The matrices are projected onto a subspace spanned by training wave
functions $\ket{\mathbf{\Psi}_{\otimes,k}} = \bigotimes_j \ket{\psi_k(\T_j)}$,
obtained by solving the forward problem at sampled
couplings~\cite{Yu:2025mep}:
\begin{multline}
    (\mathcal{K}_0)_{kl} \equiv  \braket{
      \mathbf{\Psi}_{\otimes,k} | \mathbf K_0 | \mathbf{\Psi}_{\otimes,l}
    }  \\
= \begin{vmatrix}
(V_{1}(\T_1))_{kl} & \cdots &(V_{m}(\T_1))_{kl}   \\
\vdots & \ddots & \vdots \\
(V_{1}(\T_m))_{kl} & \cdots& (V_{m}(\T_m))_{kl}
\end{vmatrix} \,,
\label{eq:projection}
\end{multline}
with $(\mathcal{K}_i)_{kl}$ obtained by replacing the $i$-th column with
$P = (P_1,\ldots,P_m)^T$, where
\begin{equation}
  (P_j)_{kl} \equiv \braket{\psi_k(\T_j) |H_{0}(\T_j)| \psi_l(\T_j)}
  - E(\T_j)\delta_{kl} \,.
\end{equation}
The reduced eigenvalue problem parameterized by truncated energies
$E(\T_j)$ then yields the couplings $C_{2n}$.
We note that the formulation above can also be trained using matrix-based emulation 
techniques based on parametric matrix models. Such approaches have been used to
approximate spectral properties and propagate theoretical uncertainties in nuclear 
systems~~\cite{Cook:2024pmm,Curry:2025qmc,Somasundaram:2024inf,Munoz:2026aem}.

Our algorithm thus proceeds in three stages. (1)~Establish the model
space using the \textit{ab initio} basis and construct the (training)
interactions within it.
(2)~Use the MEP emulator to map the truncated energies to coupling
constants. 
We find that 40 to 60 Latin Hypercube training points typically
achieve mean relative errors below $10^{-4}$ over a wide range of
energies.
This accuracy is comparable to that of standard root-finding algorithms,
but the emulator yields the \textit{entire} multivariate mapping rather
than a single data point.
(3)~Extract phase shifts and/or scattering amplitudes from the
reconstructed effective interaction using few-body methods.
In this work, we employ specifically the R-matrix method (matching to
asymptotics at a channel radius)~\cite{rmatrix} as primary tool for
step (3).
We cross-check the results against basis-specific methods employed in
the many-body context, including the spherical-wall method (imposing a
hard wall at large $R$, suitable for lattice bases)~\cite{Lu:2015riz}
and the J-Matrix/HORSE method (exploiting the effective wall created
by HO truncation)~\cite{Shirokov:2003kk} to ensure discretization
artifacts are well captured.
In our implementations, we use both a spherical HO basis (applicable
to NCSM, coupled-cluster, IMSRG, etc.) and a lattice basis (applicable
to NLEFT, used primarily with Monte Carlo methods to date, but recently
explored also with IMSRG and coupled-cluster
approaches~\cite{Rothman:2025uza}).
We furthermore note that the construction extends straightforwardly to
artificial traps~\cite{Zhang:2019cai,Zhang:2020rhz}, so overall our
MEP method is widely applicable.


\paragraph{Results.}

In the following, we test and validate our prescription using systems of
increasing complexity.
We first consider a solvable two-body model for which the inverse problem
admits a closed-form solution.
We then study neutron-alpha scattering to demonstrate applicability to
\textit{ab initio} calculations, alpha-alpha scattering to test the method
in the presence of non-perturbative Coulomb effects, and
proton-\isotope[14]O elastic scattering to establish applicability to
many-body calculations near the nuclear drip lines.

\begin{figure}[t]
    \centering
    \includegraphics[width=\linewidth]{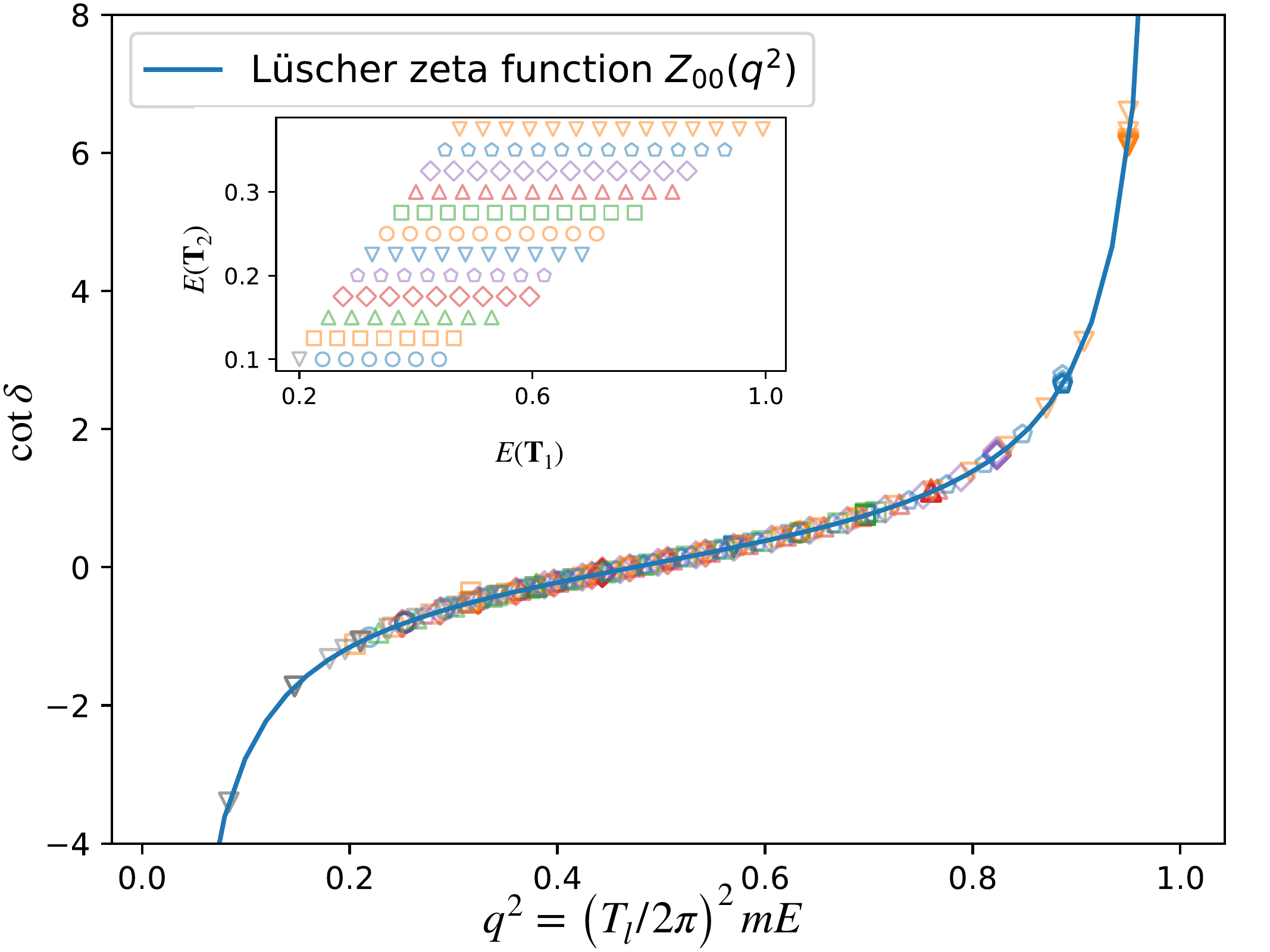}
    \caption{Toy-model benchmark of the MEP prescription with the analytical
    result~\cite{Luscher:1990ux}.
    The system is described by effective interactions with $\Lambda^{-1} = 0.8$ in
    two cubic boxes with sizes $(T_l)_1  = 6$, $(T_l)_2  = 10$, and the same
    lattice spacing $T_h = 1/6$ (in natural units).
    The MEP maps each energy pair to a set of parameters ($C_0, C_2$),
    constructing the potential at NLO for the scanned range of inputs.
    The resulting $\cot\delta$ values (via the R-matrix method) trace out the
    analytic function $Z_{00}(q^2)$ to check and visualize the extracted
    effective interactions.
    In the inset, we display truncated energy pairs scanned as input.
    Each symbol style denotes a distinct pair; the grey triangle marks all
    pairs both below (0.1, 0.2).
    The constructed interactions determine scattering at all energies up to
    their truncation scale.}
    \label{fig:twobody}
\end{figure}

\paragraph{Two-body toy model} 
As a first test, we study a case where the inverse problem has a known analytic
solution: two particles with a short-range interaction in a periodic box.
We construct a two-body Hamiltonian with the local form
Eq.~\eqref{eq:expansion-local} at $\Lambda^{-1} = 0.8$ in two box sizes, determining
the IR scales $(T_l)_1=6$ and $(T_l)_2=10$, with fixed UV truncation $T_h=1/6$.
All of these are in natural units $\hbar c = m = 1$.
As shown in Fig.~\ref{fig:twobody}, the MEP emulator maps each energy pair
(inset) to a complete set of coupling constants $(C_0, C_2)$.
The points in Fig.~\ref{fig:twobody} are from hundreds of independent inversions
and are mapped to phase shifts via the R-matrix method.
Using the dimensionless scaling of the energy $q^2 = \frac{T_l^2}{(2\pi)^2}m E$,
they collapse onto a single curve.
This common curve coincides, at every input $q^2$, with the closed-form
spectrum–amplitude relation (Lüscher formula),
$\cot\delta = Z_{00}(q^2)/(\pi^{3/2}q)$~\cite{Luscher:1986pf,Luscher:1990ux}.
This shows that the inversion is internally consistent.  We have checked that
the separable interaction~\ref{eq:expansion-sep} gives comparable results.

\begin{figure}[t]
    \centering
    \includegraphics[width=1\linewidth]{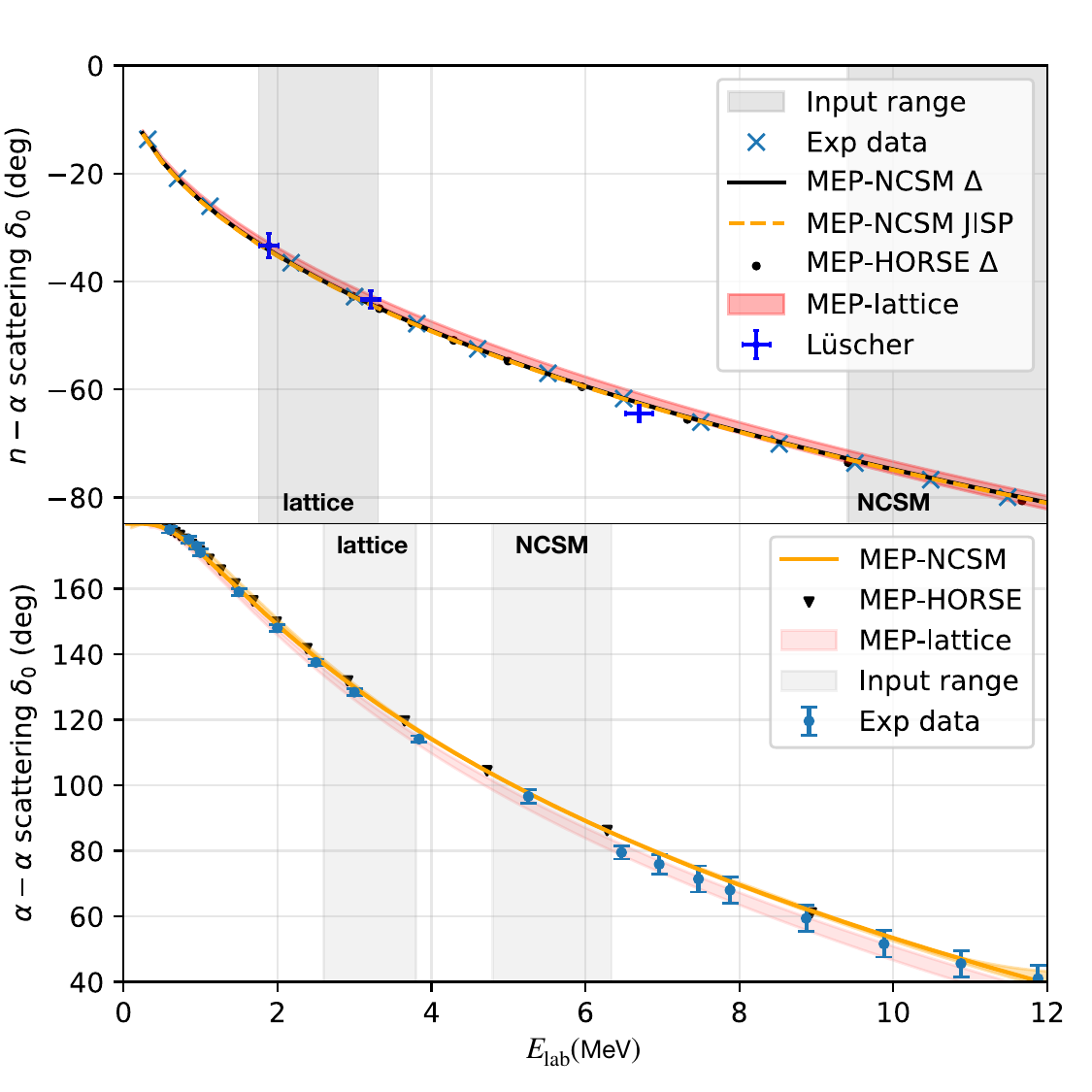}
    \caption{The top panel shows neutron-alpha $S$-wave ($^2S_{1/2}$) phase
    shifts extracted via the MEP emulator.
    Results from NCSM inputs obtained via R-matrix (MEP-NCSM) use
    $\Delta$NNLO$_{\text{GO}}$ and JISP16 (labeled as $\Delta$ and JISP,
    respectively) interactions.
    We include one HORSE validation (incorporating the discretizations artifact),
    labeled MEP-HORSE, with the $\Delta$NNLO$_{\text{GO}}$.
    We also present NLEFT lattice results (MEP-lattice), compared against the
    standard L\"{u}scher method. The shaded band reflects propagated lattice
    energy uncertainties including the choices of cutoffs.
    The bottom panel shows alpha-alpha $S$-wave phase shifts.
    The MEP emulator reconstructs the resonant behavior from NCSM truncated
    energies (using JISP16) and NLEFT lattice results, handling the Coulomb
    interaction non-perturbatively.
    Results agree with experimental data (blue error
    bars)~\cite{Elhatisari:2025fyu, AFZAL:1969iik}.
    The shaded vertical band indicates the NCSM training range.
    MEP uncertainties are assigned by varying the input truncations and
    operator expansions.}
    \label{fig:alpha}
\end{figure}

\paragraph{Neutron-alpha scattering}
Having validated the prescription against the analytic two-body result, we
apply the MEP prescription to neutron-alpha scattering in the $^2S_{1/2}$
channel, one of the simplest nuclear cluster systems and a standard benchmark
for nuclear continuum calculations~\cite{Kravvaris:2020lhp,Burrows:2023ygq,%
Yang:2025mhg,Shirokov:2018nlj,Elhatisari:2025fyu}.
$S$-wave scattering in this system does not depend on the less well-known
chiral three-body forces, making it ideal for benchmarking methods.
We analyze NLEFT lattice spectra~\cite{Elhatisari:2025fyu} at box sizes $L=9.2~\text{fm}$ 
and $L=10.6~\text{fm}$ with N3LO chiral interactions, as well as NCSM spectra~\cite{Bigstick}
at $N_{\text{max}}=9,11$ with oscillator frequency $\hbar\Omega = 30$~MeV,
generated by both the JISP16 interaction~\cite{Maris:2008ax,Maris:2013poa} and
the nucleon-nucleon part of the $\Delta$NNLO$_{\rm GO}$ interaction (with 394
MeV cutoff)~\cite{Ekstrom:2017koy,Jiang:2020the}.
As shown in the top half of Fig.~\ref{fig:alpha}, the same MEP prescription
applied to different \textit{ab initio} methods and interactions produces
consistent phase shifts that are also in excellent agreement with experimental
data.
Moreover, again we find agreement with results obtained directly from the
Lüscher formalism.

\paragraph{Alpha-alpha scattering}
The $\alpha$-$\alpha$ system (hosting the $^8$Be resonance) provides a more
stringent test for truncated-space calculations~\cite{Elhatisari:2021eyg,%
Kravvaris:2020cvn}.
With a Sommerfeld parameter $\eta = \frac{ \mu \alpha_{\text{EM}} Z_1 Z_2} 
{k} \sim 1$ (where $\mu$ the reduced mass, $\alpha_{\text{EM}}$ the
fine-structure constant, $Z_1,Z_2$ the charges of two clusters, and $k$ the
relative momentum) at a relative energy of 1~MeV, the strong Coulomb repulsion
demands a non-perturbative treatment, and no L\"{u}scher-type quantization
condition currently exists for the scattering of charged particles on a
periodic lattice (known relations are either limited to perturbative Coulomb
effects~\cite{Beane:2014qha,Stellin:2020gst} or to bound
states~\cite{Yu:2022nzm}).
The system also features a narrow $0^+$ ground-state resonance, yet within reach of exact NCSM calculations.
Our MEP prescription handles the Coulomb interaction by incorporating it
directly into $H_0$. 

In the bottom half of Fig.~\ref{fig:alpha}, the MEP emulator, trained
on the convergence patterns of the NCSM and lattice results, captures
the full resonant behavior near threshold, and a very shallow resonance
around 50--100~keV is identified. 
For the NCSM calculations, we use the JISP16 interaction
($N_{\text{max}}=10,12$ with oscillator frequency $\hbar\Omega = 25$~MeV),
which effectively incorporates three-body force contributions, to
demonstrate the method's capability in the Coulomb regime.
For lattice calculation the same N3LO chiral interaction has been used,
with box sizes $L=13.2~ \text{fm}$ and $L=15.8~ \text{fm}$.
We note that the extension to lattice-based $\alpha$-$\alpha$ calculations
with non-perturbative Coulomb can introduce additional subtleties, including
topological factors~\cite{Bour:2011top}, and we will present details in
a forthcoming publication.

\begin{figure}[t]
    \centering
    \includegraphics[trim={5mm 0mm 10mm 5mm}, clip, width=1\linewidth]{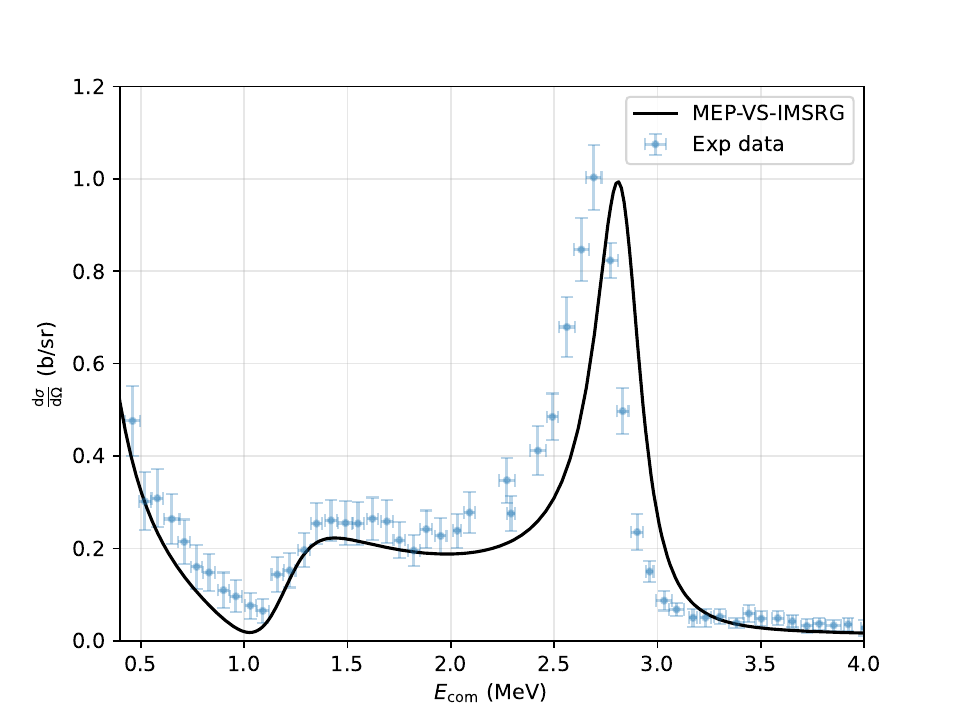}
    \caption{Differential cross section at $180^\circ$ for
    proton-\isotope[14]O elastic scattering, comparing the MEP prediction from
    VS-IMSRG calculations with experimental data~\cite{DeGrancey:2016bez}.
    The Hamiltonian is evolved into thevalence space with an \isotope[14]O core.}
    \label{fig:po14}
\end{figure}

\begin{table}[htbp]
\centering
\caption{Comparison of experimental and theoretical resonance predictions of
\isotope[15]F.
We use the Breit-Wigner form to fit the phase shifts obtained from the MEP
method.
The reported experimental errors are (statistical) and (systematic) errors.
In our prediction the errors are (interaction) and (method) errors.
Interaction errors include different cutoffs, truncations, and effective
interaction expansions including the subsequent Gaussian term $C_4$.
We assign the method error to the variations in calculating relative energy.}
\label{tab:resonances}
\begin{tabular}{c c c c c}
\hline\hline
$J^\pi$ & $E_R^{\mathrm{exp}}$ (MeV) & $\Gamma^{\mathrm{exp}}$ (keV)
& $E_R^{\text{MEP}}$ (MeV) & $\Gamma^{\text{MEP}}$ (keV) \\
\hline
$1/2^+$ & 1.27(2)(2) & 374(70)(+200) & 1.25(3)(10) & 379(14)(50) \\
$5/2^+$ & 2.81(12) & 251(26) & 2.83(5)(20) & 265(20)(50)  \\
\hline\hline
\end{tabular}
\end{table}

\paragraph{Proton-\isotope[14]O scattering}
Having validated the prescription on light systems, we demonstrate its
portability to an approximate many-body framework by applying it
to Valence-Space (VS) IMSRG~\cite{Hergert:2020bxy,Stroberg:2019mxo}
calculations of proton-\isotope[14]O scattering.
Here we employ the same $\Delta$NNLO$_{\rm GO}$ interaction~\cite{Jiang:2020the}
with three-nucleon forces truncated via the normal-order two-body approximation
at $E_{3,\text{max}} = 24, 28$~\cite{Miyagi:2023qce}.
The basis itself is truncated at $e_{\text{max}} = 10$--$14$ (single-particle
basis truncation) with $\hbar\Omega = 16, 18$~MeV.
The IMSRG evolves the Hamiltonian into the proton $sd$-shell valence space with
an \isotope[14]O core, from which we compute the truncation-dependent
relative energy $E_{\text{rel}}(\T)= \braket{\psi_{j}|H(\T) - a_j^\dagger
H(\T) a_j|\psi_{j}}$,
the operator form of $E_{\isotope[15]F,j+}(\T) - E_{\isotope[14]O,0+}(\T)$
(with $j$ the total angular momentum) evaluated with the \isotope[15]{F}
state $\ket{\psi_{j}}$.
The $\T$ here denotes truncations by $e_{\text{max}}$ and $\hbar \Omega$.
Two resonances, $1/2^+$ and $5/2^+$, are successfully identified in the
$sd$ valence space (Table~\ref{tab:resonances}), in good agreement with
experimental values~\cite{DeGrancey:2016bez,Girard-Alcindor:2021xgs}.
The first quoted uncertainties reflect different selections of
$e_{\text{max}}$, variations of the interaction cutoff scale from 50 to 130~MeV,
and from including an additional coupling $C_4$ (multiplying a
four-derivative regulated contact operator), to assess the model
independence of the extracted resonance parameters.
The second quoted uncertainties reflect the impact of the reference states
used to compute the relative energy.
We note that studies resonances with energies above 4~MeV require 
including single-particle excitations beyond the present valence space; this
is therefore left to future work.
In addition, Fig.~\ref{fig:po14} shows the predicted differential cross section
at $180^\circ$ (center-of-mass), compared to experimental data.
Overall, this application demonstrates that the MEP prescription ports directly
to valence-space methods -- with no modification to the many-body
solver required because only the truncation-dependent energies are
needed as input.


\paragraph{Conclusion and Outlook.}
The connection between discrete spectra from truncated-basis calculations
and effective few-body cluster interactions suggests that the natural
objects to extract from the discrete input can be the coupling constants
parameterizing an effective interaction.
In this work, we provide an MEP based approach that can carry out the
mapping from spectra to effective interactions in a highly
efficient manner.

The MEP emulator yields a general prescription from which all scattering
observables can be obtained straightforwardly, regardless of which
underlying basis truncation or \textit{ab initio} technique was employed
on the many-body side.
In the context of finite-volume simulations, the MEP method provides an
alternative to the well-established L\"{u}scher formalism
that avoids several challanges by working on the level of the potential
instead of the scattering amplitude.
Because of this, the MEP also naturally handles non-perturbative
Coulomb effects and coupled channels, which require intricate analytic
extensions in amplitude-level approaches.
We also expect this method should work in more general setups beyond
the low-energy nuclear physics.
For example, lattice gauge theories such as Lattice QCD+QED dynamically
generate electromagnetic effects through fluctuating $U(1)$ gauge
fields.
This is different from \textit{ab initio} frameworks where the
proton-proton electromagnetic interaction is an input feature of the
Hamiltonian.
A key strength of the MEP approach is that it is agnostic to these
differences.
Because the MEP emulator operates as a downstream inversion tool on the
resulting truncated energy spectra, it can extract the parameters for
the corresponding effective theory independent of the microscopic origins.

We demonstrated the performance of our approach with three
\textit{ab initio} frameworks (NLEFT, NCSM, VS-IMSRG), culminating in
a prediction of proton-\isotope[14]O resonances from a numerical
calculation of only deeply bound states.
No input is required beyond the standard many-body calculation of
energies at different basis truncations, a variation that is routinely
used to test convergence for bound states~\cite{Furnstahl:2013vda}.

Looking ahead, one next step for the approach presented in this Letter
is to incorporate the uncertainty quantification (UQ) framework already
established for MEP emulators in general~\cite{Yu:2025mep}.
Proper UQ will be particularly important for constraining three-body
interactions for systematically understanding the role of such forces
in the continuum.
We also plan to find consistent schemes for computing truncated
relative energies in methods with reference-state dependence, such as
coupled-cluster and IMSRG.
Importantly, while the present approach focused on constraining generic
effective interactions, providing an important proof-of-principle
demonstration, in the future we plan to apply MEP emulation to
constrain the LECs of genuine EFT interactions, in a rigorously
renormalized order-by-order approach.
Consistently mapping the \textit{ab initio} input, which may itself
come with an expansion at the microscopic side, to the power-counting
in a few-body Halo/Cluster EFT is the key challenge that needs to
be addressed for implementing such a program.

\begin{acknowledgments}
We thank Avik Sarkar, Congwu Wang, Matthias Heinz, and Noritaka Shimizu
for helpful discussions.
H.Y.\ and T.M.\ are in part supported by JST ERATO Grant No.~JPMJER2304,
Japan.
T.M.\ is also in part supported by JSPS KAKENHI Grant No.~25K07294, 25K00995, 25K07330, and~26H01394.
H.Y.\ and T.M.\ are also in part supported by the Multidisciplinary
Cooperative Research Program in CCS, University of Tsukuba. 
D.L.\ and Y.M.\ are supported by the U.S.\ Department of Energy Grants
DE-S0013365, DE-SC0023175, DE-SC0026198, and DE-SC0023658.
The work of S.K. was supported in part by the U.S.\ National Science
Foundation under Grant No.~PHY--2044632
as well as by the U.S.\ Department of Energy under Grant No.~DE-SC0024520 and~DE-SC0026198.
The work of S.E. is supported in part by the Scientific and Technological Research Council
of Turkey (TUBITAK project no. 123F464). 
The VS-IMSRG calculations were performed using the imsrg++~\cite{imsrg}
and KSHELL~\cite{kshell} codes.
The NCSM calculations were performed using the BIGSTICK
code~\cite{johnson_bigstick}.
\end{acknowledgments}

\bibliography{refs}

\end{document}